\documentclass[aps,prb,twocolumn,twoside,superscriptaddress]{revtex4}
\usepackage[version=3]{mhchem} 
\usepackage{amsmath}
\usepackage{amssymb}
\usepackage{amsthm}
\usepackage{amsfonts}
\usepackage{listings}
\usepackage{longtable}
\usepackage{enumerate}
\usepackage{latexsym}
\usepackage{color}
\usepackage{setspace} 
\usepackage{blindtext}
\usepackage{dsfont}
\usepackage{mathrsfs}
\usepackage{array,etoolbox}
\usepackage{bm}
\usepackage{hyperref}
\usepackage{graphicx}
\usepackage{subfigure}
\usepackage{psfrag}
\usepackage{changes}
\usepackage{comment}

\usepackage{physics}
\usepackage{graphicx}
\usepackage{amsmath}
\usepackage{float}
\usepackage{amssymb}
\usepackage{placeins}
\usepackage{array}

\newcommand{\vecauto}[1]{\ensuremath{\vectorbold*{#1}}}
\newcommand{\vQ}{\vecauto{Q}_{\textrm{ex}}}

\newcommand{\vq}{\vecauto{q}_{\textrm{L}}}

\newcommand{\Hint}{H^{\textrm{int}}}

\newcommand{\QEx}{Q_{\textrm{ex}}}

\begin{document}

\newcommand{\TODO}[1]{\textcolor{red}{#1}}
\newcommand{\super}[1]{\ensuremath{^{\mathrm{#1}}}}

\graphicspath{{images/}{./images/}}

\title{Substrate Effect on Excitonic Shift and Radiative Lifetime of Two-Dimensional Materials}

\setcounter{page}{1}
  \date{\today}

\author{Chunhao Guo}
\affiliation{%
 Department of Chemistry and Biochemistry, University of California Santa Cruz, Santa Cruz, CA, 95064, USA
}%
\author{Junqing Xu}
\affiliation{%
 Department of Chemistry and Biochemistry, University of California Santa Cruz, Santa Cruz, CA, 95064, USA
}%

\author{Yuan Ping}
\email{yuanping@ucsc.edu}
\affiliation{%
 Department of Chemistry and Biochemistry, University of California Santa Cruz, Santa Cruz, CA, 95064, USA
}%
\date{\today}

\begin{abstract}

Substrates have strong effects on optoelectronic properties of two-dimensional (2D) materials, which have  emerged as promising platforms for exotic physical phenomena and outstanding applications. To reliably interpret experimental results and predict such effects at 2D interfaces, theoretical methods accurately describing electron correlation and electron-hole interaction such as first-principles many-body perturbation theory are necessary. 
In our previous work [Phys. Rev. B 102, 205113(2020)], we developed the reciprocal-space linear interpolation method that can take into account the effects of substrate screening for arbitrarily lattice-mismatched interfaces at the GW level of approximation. In this work, we apply this method to examine the substrate effect on excitonic excitation and recombination of 2D materials by solving the Bethe-Salpeter equation. We predict the nonrigid shift of 1s and 2s excitonic peaks due to substrate screening, in excellent agreements with experiments. We then reveal its underlying physical mechanism through 2D hydrogen model and the linear relation between quasiparticle gaps and exciton binding energies when varying the substrate screening.
At the end, we calculate the exciton radiative lifetime of monolayer hexagonal boron nitride with various substrates at zero and room temperature, as well as the one of WS$_2$ where we obtain good agreement with
experimental lifetime. Our work answers important questions of substrate effects on excitonic properties of 2D interfaces.

\end{abstract}

\maketitle


\section{Introduction}

Due to reduced dimensionality, two-dimensional (2D) materials and their heterostructures have shown emerging optical properties, such as strong light-matter interaction and giant excitonic binding energy \cite{chernikov2014exciton, qiu2013optical}, distinct from the three-dimensional counterparts. Promising applications have been demonstrated in many areas, such as opto-spintronic devices~\cite{wolf2001spintronics, vzutic2004spintronics} and quantum information technologies~\cite{he2015single, tran2016quantum}. 
Experimentally, growth of 2D materials, achieved through physical epitaxy or chemical vapor deposition (CVD), is typically supported on a substrate~\cite{novoselov20162d}. Similarly, the optical measurements, such as photoluminescence and absorption spectra, are often performed on top of substrates or sandwiched by supporting substrates. 
In general, the optoelectronic properties of 2D materials can be strongly modified by environmental dielectric screening. For example, their fundamental electronic gap and exciton binding energy can be significantly reduced at presence of substrates when forming heterostructures~\cite{winther2017band,ugeda2014giant}. 


An interesting experimental observation is that in the presence of environmental dielectric screening (including increasing the number of layers of 2D materials), the 2s (second) exciton peaks shift strongly, but the 1s (first) exciton peaks stay relatively unchanged
~\cite{chernikov2014exciton,ugeda2014giant,waldecker2019rigid,raja2017coulomb}.
Yet, the physical origin of such non-rigid shift of excitonic peaks due to substrate screening has not been revealed and requires careful investigation.
Its quantitative prediction is also crucial for correct interpretation and utilization of experimental measurement data.
For example, 
the energy difference between $1s$ and $2s$ absorption peaks $\Delta_{12}$ in the presence of different substrate screening has been used to estimate electronic band gaps in optical measurements  ~\cite{raja2017coulomb,waldecker2019rigid}, although its underlying assumption still requires careful justification. 

Physically, the exciton peak shift due to substrate screening is determined by changes both from the electronic gap and exciton binding energy, which compete with each other. Therefore, theoretical methods such as many-body perturbation theory (MBPT, GW approximation and solving Bethe-Salpeter equation (BSE)) including accurate electron correlation and electron-hole interactions are necessary to accurately describe both electronic gaps and exciton binding energies~\cite{PingCSR2013}. In order to study the effect of various substrates at such level of theory, our recent development on substrate dielectric screening from MBPT~\cite{guo2020substrate} will make these calculations computationally tractable. There we developed a reciprocal linear-interpolation method, which interpolates the dielectric matrix elements from substrates to materials at the entire $\vec{q}+\vec{G}$ space, thus completely removes the constraint on symmetry and lattice parameters of two interface systems. In this work we will further apply this method to study the substrate effects on excitonic excitation energies and radiative lifetime. 

Previous theoretical studies well described the exciton energy spectrum of free-standing 2D materials with relative simple models, e.g. 2D Wannier exciton Rydberg series~\cite{olsen2016simple} or linear scaling between exciton binding energy and electronic band gap~\cite{choi2015linear, jiang2017scaling}. The environmental screening induced exciton peak shifts have been discussed with semi-infinite dielectric models~\cite{cho2018environmentally}.
The applicability of these models to explain the excitonic physics of 2D heterostructures or multilayer systems is unclear and requires examination. 
On the other hand, past first-principle work mostly focused on the substrate effects on first exciton peak position (optical gap) and electronic band gaps~\cite{ugeda2014giant}. The key question of the origin of nonrigid shift of excitonic peaks due to substrate screening and whether there is any universal scaling relation have not been answered to our best knowledge. 
Furthermore, how the substrates affect exciton radiative lifetime, a critical parameter determining quantum efficiency in optoelectronic applications, has rarely been studied before.
Understanding how radiative lifetime changes in the presence of substrate screening will provide important insights to experimental design of optimal 2D interfaces.

In this paper, we will answer the outstanding questions of the substrate effect on exciton excitation and radiative lifetime at 2D interfaces, through first-principles many-body perturbation theory. 
We reveal the origin of experimentally observed nonrigid shift of $1s$ and $2s$ exciton peaks and their scaling universality induced by substrate screening~\cite{chernikov2014exciton}. 
We then demonstrate the relation between exciton binding energy and electronic band gap due to substrate effects~\cite{choi2015linear, jiang2017scaling} in comparison with
the case of free-standing 2D materials.
At the end, we elucidate the effect of substrate screening on the exciton lifetime of 2D materials at both zero and finite temperature. 

\section{Computational Details}

The ground state calculations are performed based on Density functional theory (DFT) with the Perdew-Burke-Ernzerhof (PBE) exchange-correlation functional~\cite{perdew1996generalized}, using the open source plane-wave code QuantumEspresso~\cite{Giannozzi_2017}. We used Optimized Norm-Conserving Vanderbilt (ONCV) pseudopotentials~\cite{hamann2013optimized} and a 80 Ry wave function cutoff for most systems except WS$_2$ (60 Ry). For monolayer WS$_2$, spin-orbit coupling is
included through fully relativistic ONCV pseudopotentials. 
The interlayer distance and lattice constants of hBN interfaces and multiplayer WS$_2$ are obtained
with PBE functionals with Van der Waals corrections~\cite{grimme2006semiempirical, barone2009role}. 

In this paper, the quasiparticle energies and optical properties are calculated with many-body perturbation theory at GW approximation and solving Bethe-Salpeter equation respectively, for hBN/substrate interfaces and multi-layer WS$_2$. To take into account the effect of substrates, we use our recently developed sum-up effective polarizability  approach ($\chi_{\text{eff}}$-sum)~\cite{guo2020substrate}, implemented in a postprocessing code interfacing  with the Yambo-code~\cite{marini2009yambo}. Briefly, we separate the total interface systems into subsystems~\cite{yan2011nonlocal} and perform GW/BSE calculations for monolayer hBN or WS$_2$ including the environmental screenings by the $\chi_{\text{eff}}$-sum method.
For lattice-mismatched interfaces, we use our reciprocal-space linear interpolation technique~\cite{guo2020substrate} to interpolate the corresponding matrix elements from substrate to materials $\vec{q}+\vec{G}$ space before summing up the subsystems' effective polarizabilities.

{In order to speed up convergence with respect to vacuum sizes, a 2D Coulomb truncation technique~\cite{RozziPRB2006} was applied to GW and BSE calculations.
The k-point convergence of quasiparticle gaps and BSE spectra for monolayer hBN is shown in SI Figure 2 and 3, where we show $36\times36\times1$ k points converge up to 20 meV, which was adopted for other calculations.}
More details of interface structural parameters and GW/BSE convergence tests can be found in Supporting Information.  

\section{Results and discussions}

\subsection{Substrate screening effect on optical excitation energy of hBN}
Hexagonal boron nitride (hBN) has drawn significant attentions recently due to its potentials as host materials for single photon emitters~\cite{Mendelson2020wc,Tran2016vw} and spin qubits~\cite{Gottscholl2020vz,Turiansky2020ww} for quantum information science applications. Rapid progress has been made both experimentally and theoretically~\cite{wu2019carrier,smart2018fundamental,wu2017first,smart2020intersystem,Sajid_2020}. The related optical measurements are often performed on top of substrates, whose effects have not been carefully examined. 
We use hBN as a prototypical example to examine how optical excitation energies are changed in the presence of substrates. We obtain the optical excitation energies and absorption spectra by solving the BSE (with electron-hole) and Random Phase approximation (RPA) calculations (without electron-hole interactions),  with GW quasiparticle energies as input.

\begin{figure}
    \includegraphics[width=0.45\textwidth]{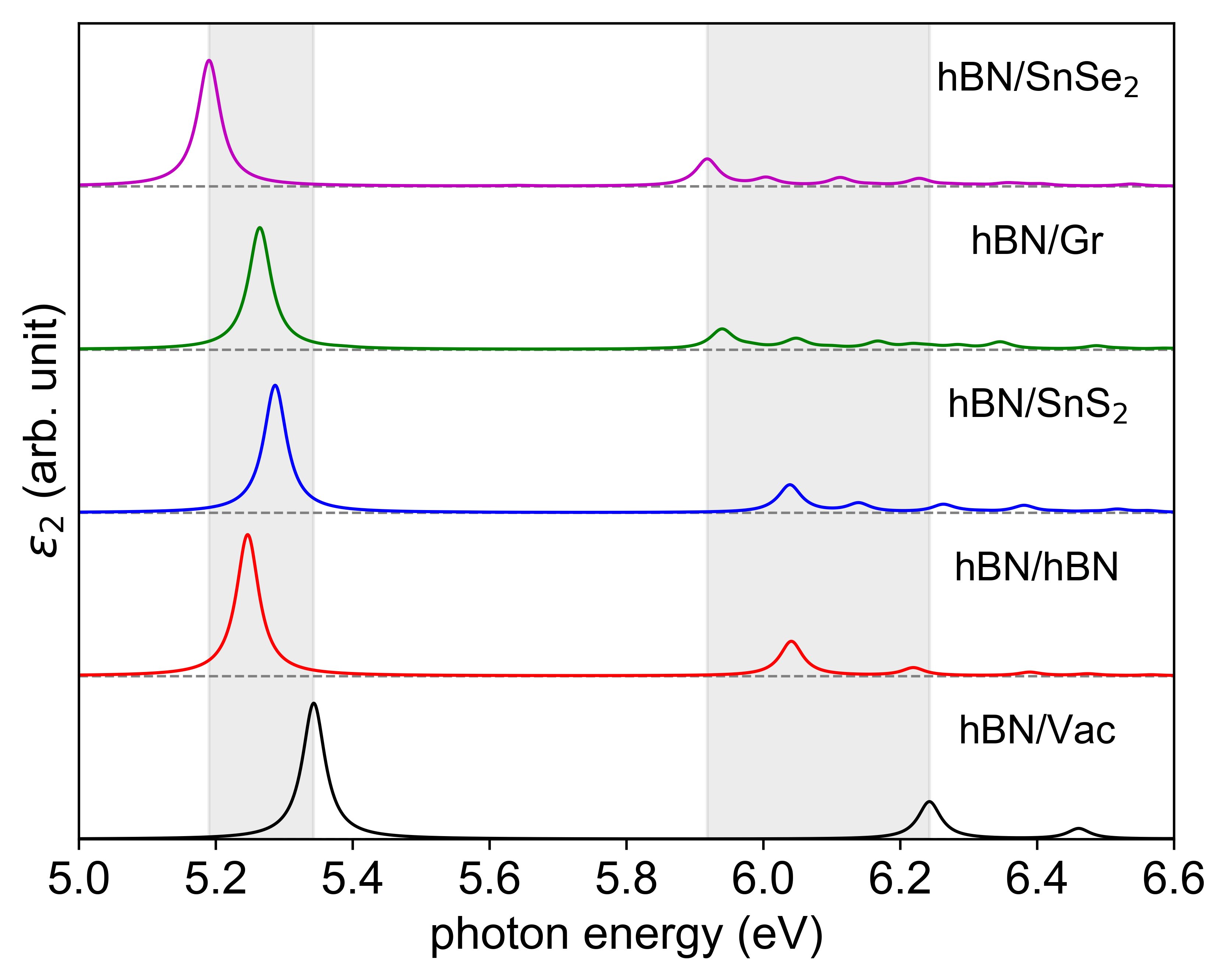}
    \centering
  \caption{\label{BSE1} Absorption spectra obtained by solving BSE (with e-h interaction) for monolayer hBN interfacing with various substrates. The curves from the bottom to the top are for 1) free-standing hBN (hBN/Vac) 2) hBN/hBN 3) hBN/SnS$_2$ 4) hBN/Graphene(Gr) 5) hBN/SnSe$_2$ heterostructures. Curves are vertically displaced for clarity.}
\end{figure}

Figure~\ref{BSE1} shows the BSE calculations of monolayer hBN with various substrates, including SnSe$_2$, graphene, SnS$_2$, hBN as well as without substrate (interfacing with vacuum). The 1s absorption peak shifted little referenced to the free-standing hBN (black curve) i.e. $<$ 0.2 eV but the 2s absorption peak shifts nearly twice compared to the first peak. This trend obtained from our BSE calculations is fully consistent with the experimental observations mentioned in the introduction~\cite{chernikov2014exciton}.

\begin{figure}
    \includegraphics[width=0.45\textwidth]{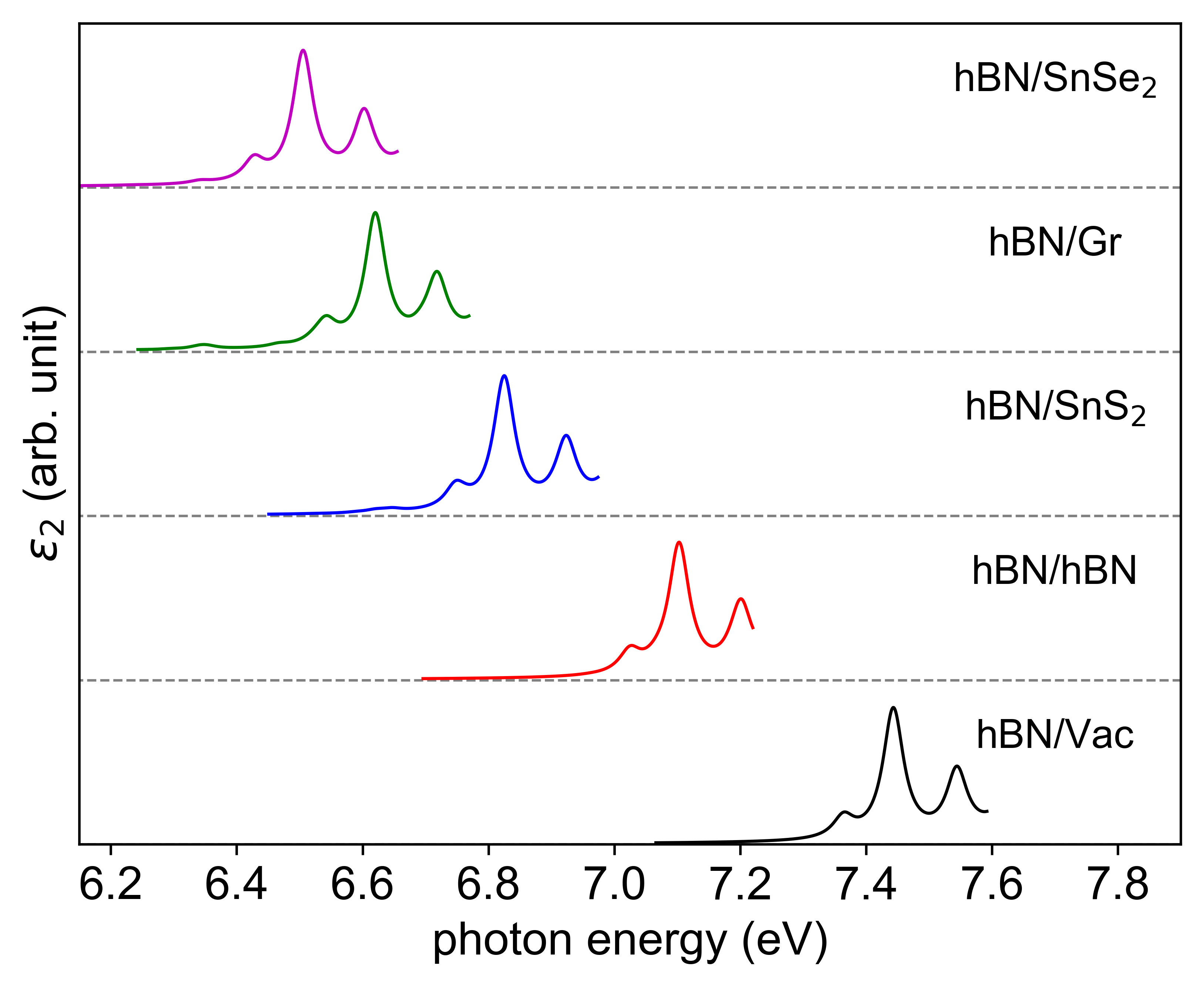}
    \centering
  \caption{\label{RPA} Absorption spectra at RPA with GW quasiparticle energies (without e-h interaction). The curves from the bottom to the top are for 1) free-standing hBN (hBN/Vac) 2) hBN/hBN 3) hBN/SnS$_2$ 4) hBN/Graphene(Gr) 5) hBN/SnSe$_2$ heterostructures. Curves are vertically displaced for clarity.}
\end{figure}

In contrast, Figure~\ref{RPA} shows the RPA spectra of hBN with various substrates using GW quasiparticle energies exhibit a nearly-rigid shift to a lower energy (compared to free-standing hBN). The red shift is mainly due to the reduction of electronic band gap in the presence of substrate screening. This rigid shift may be qualitatively explained by the independence of k-point for electronic band structure under Born approximation \cite{cho2018environmentally, waldecker2019rigid}, which has been reported for 2D semiconductors (e.g. WS$_2$) \cite{waldecker2019rigid}. 

\begin{figure}
    \includegraphics[width=0.5\textwidth]{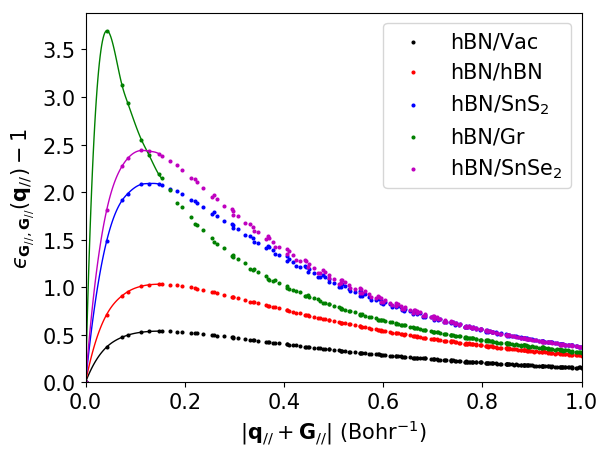}
    \centering
  \caption{\label{eps} The in-plane diagonal elements of RPA dielectric matrix $\epsilon_{\mathbf{G}_{||}, \mathbf{G}_{||}}(\mathbf{q}_{||})$-1 as a function of absolute values of in-plane momentum transfer $\mathbf{q}_{||}+\mathbf{G}_{||}$ for different hBN/substrate interfaces.}
\end{figure}

{In general, we find the reduction of quasiparticle band gaps due to substrates increases with stronger substrate dielectric screening. 
However, a simple dielectric constant picture is insufficient to describe low dimensional systems. 
Specifically, we show the in-plane diagonal elements of dielectric matrices in Figure~\ref{eps}.
For example, comparing with the SnSe$_2$ substrate (purple dots), the graphene substrate (green dots) has a stronger dielectric screening at a small momentum transfer region close to zero, and a weaker dielectric screening at a larger momentum transfer region. 
As a result, the reduction of band gap with the SnSe$_2$ substrate is larger than the one with the graphene substrate although graphene is closer to a metallic system at the Dirac cone than SnSe$_2$.
Therefore, fully first principle calculations are required to get reliable prediction of screening effects by various substrates.}

\subsection{1s and 2s exciton binding energy change with substrate screening} \label{scal-ebeb}


The difference between BSE excitation energies ($E_{\text{S}}$) and electronic band gaps $E_g$ defines exciton binding energy $E_b$ for excitonic state $S$:
\begin{align}
    E_b (S) = E_g - E_{\text{S}}.
\end{align}

\begin{figure}
    \includegraphics[width=0.45 \textwidth]{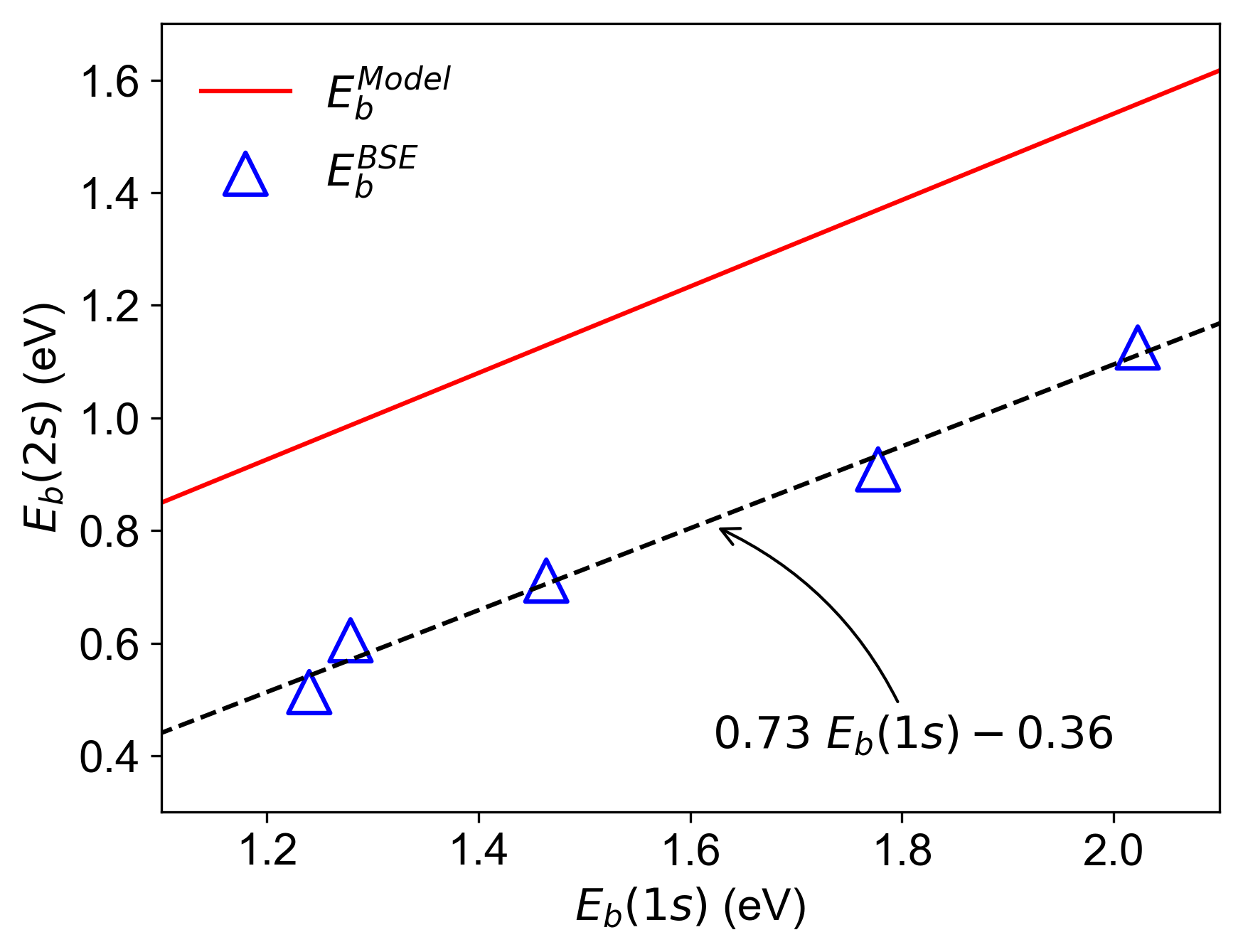}
    \centering

\caption{\label{figNonrigid} Linear scaling between exciton binding energy of 1s state $E_b(1s)$ and 2s state $E_b(2s)$. The blue triangles are first-principles results by solving BSE. The dashed black line is the linear fit to the blue triangles. The red line is computed from the 2D hydrogen model for exciton binding energies~\cite{olsen2016simple} based on Eq.~\ref{eq_lin_eb}.
The blue triangle points from right to left are 1) free-standing hBN (hBN/Vac) 2) hBN/hBN 3) hBN/SnS$_2$ 4) hBN/Graphene(Gr) 5) hBN/SnSe$_2$ heterostructures.} 
\end{figure}

We found the proportionality between 1s and 2s exciton binding energies across different substrates falls into a linear relation (i.e. with a slope of 0.73 for $E_b(2s)/E_b(1s)$), as shown in Figure~\ref{figNonrigid}.

To understand the physical meaning of this linear scaling obtained by solving BSE, we compare our results with the previous 2D hydrogen model of excitons~\cite{chernikov2014exciton, olsen2016simple}, which has been used to interpret the exciton energies of free-standing 2D materials. Here we will test the applicability of this model for substrate screening effect on 2D excitons.
In this model, we express the 2D dielectric function $\epsilon(\mathbf{q})$ as $\epsilon(\mathbf{q}) = 1 + 2 \pi \alpha {\mathbf{q}}$,  where $\alpha$ is the 2D polarizability. The exciton binding energies of $n$th 2D Rydberg-like excitonic state~\cite{yang1991analytic} ($E^{\text{Model}}_b(n)$) can be expressed as: 

\begin{equation} \label{eq_eb}
    E^{\text{Model}}_b(n) = \frac{\mu}{2 (n- \frac{1}{2})^2 \epsilon^2_n},
\end{equation}
where $\mu$ is the exciton reduced mass and $\epsilon_n$ is the effective dielectric constant for $n\text{th}$ excitonic state, defined as~\cite{olsen2016simple}:
\begin{equation} \label{e_n}
   \epsilon_n=\frac{1}{2}(1 + \sqrt{1+\frac{32\pi\alpha\mu}{9n(n-1)+3}}).
\end{equation}
Further simplification~\cite{olsen2016simple, jiang2017scaling} of Eq.~\ref{eq_eb} with Eq.~\ref{e_n} gives 2D exciton binding energy $E^{\text{Model}}_b$ independent of exciton reduced mass $\mu$ as follows:

\begin{equation} \label{eq_lin_eb}
    E^{\text{Model}}_b(n) \approx \frac{9n(n-1) +3}{16 \pi (n- \frac{1}{2})^2} \cdot \frac{1}{\alpha}.
\end{equation}


From Eq.~\ref{eq_lin_eb}, we have $E^\text{Model}_b(1s) \approx 3/(4\pi\alpha)$ and $E^\text{Model}_b(2s) \approx 7/(12\pi\alpha)$. The ratio between $E_b(2s)$ and $E_b(1s)$ is a constant $0.78$ from this simplified model. Figure~\ref{figNonrigid} shows the linear scaling between 1s and 2s exciton binding energies of monolayer hBN when changing its substrates.
The scaling behavior by the model in Eq.~\ref{eq_lin_eb} (red curve) is in qualitative agreement with our GW/BSE results (blue triangles), i.e. both of which have a linear relation between 2s and 1s exciton binding energies, with a ratio of less than one (Model: 0.78; GW/BSE: 0.73), corresponding to the slope. 
This implies the change of 1s $E_b$ is larger than the change of 2s $E_b$ due to substrate screening, with a constant ratio while varying the substrate materials. 
Figure~\ref{figNonrigid} also shows a constant shift between the first-principles scaling and the model one. This discrepancy independent of specific screening environment may come from the limitation of 2D hydrogen model, i.e. either the strict 2D limit of $\epsilon(\mathbf{q})$ is unrealistic considering the finite thickness of materials, or hBN has tighter-bounded excitons~\cite{galvani2016excitons}, deviated from 2D Wannier excitons assumed in previous 2D hydrogen model. Note that the explicit form of Eq.~\ref{eq_lin_eb} and related linear scaling behavior is based on strict 2D limit of Eq.~\ref{eq_eb}, which is better to describe the interface with relatively small thickness, e.g. heterostructures formed by atomically thin materials. We anticipate that this simplified model may become inappropriate for interface systems with large thickness, especially for systems with high dielectric semi-infinite substrates ~\cite{riis2020anomalous}.

\subsection{Substrate induced linear scaling relation between ${E_b}$ and $E_g$} \label{scal-ebeg}

The exciton peak positions are determined by both the exciton binding energies and electronic band gaps, which have opposite trends while increasing substrate screenings. 
The relationship between these two quantities was studied for free-standing 2D systems~\cite{choi2015linear, jiang2017scaling}, where $E_b\approx \frac{1}{4}E_g$ across a wide range of 2D materials.Yet, no investigation on their relationship when varying substrates has been carried out.  

\begin{figure}
    \includegraphics[width=0.45\textwidth]{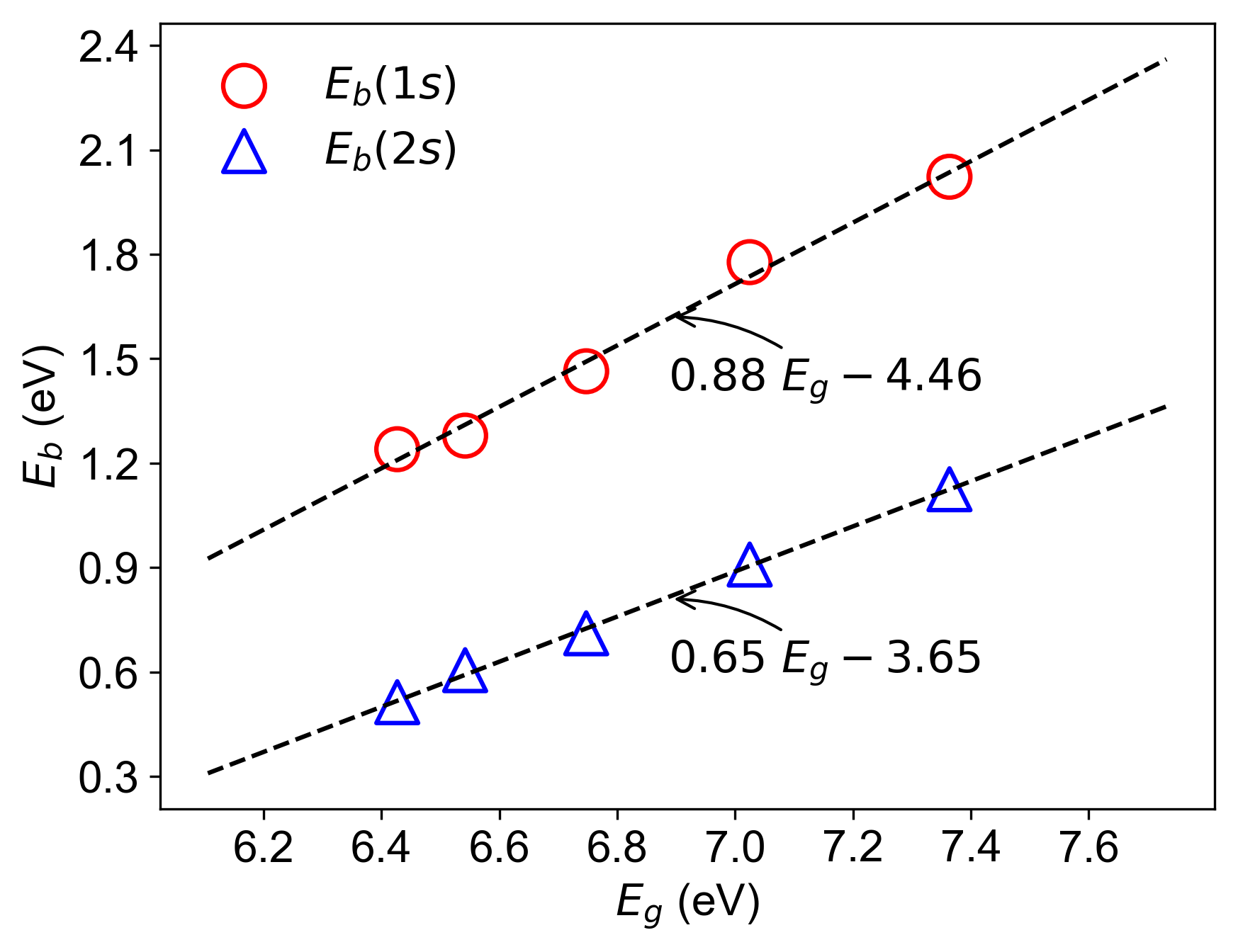}
    \centering

  \caption{\label{linear-Eg} Linear scaling between 1s and 2s exciton binding energy $E_b$ and quasiparticle direct band gaps $E_g$ computed from GW and BSE. 
  The data points from right to left are (1) free-standing hBN, (2)(2L-AA$^\prime$) hBN/hBN (3) hBN/SnS$_2$, (4) hBN/Graphene(Gr), (5) hBN/SnSe$_2$ results with effective polarizability approach ('$\chi_{\text{eff}}$-sum' method).
  } 
\end{figure}

In Figure~\ref{linear-Eg}, we show the calculated electronic gap ($E_g$) and exciton binding energy of 1s (red circle) and 2s (blue triangle) ($E_b$) for monolayer hBN at various substrates. 
Our first-principle results show a linear scaling between exciton binding energy $E_b$ (1s peak position) and quasiparticle electronic band gap $E_g$ due to substrate screening, with a slope nearly close to one (0.88 in Figure~\ref{linear-Eg}, linear fitting the computed data (red circles) with a black dashed line). This indicates that the changes of 1s exciton binding energy $\Delta E_b$ and electronic gaps $\Delta E_g$ due to substrate screening are largely canceled out. Therefore, the first exciton peak (1s) is at a relatively stable position, insensitive to the environmental screening. This explains the experimental and theoretical results in Figures~\ref{BSE1} and ~\ref{figBSE1}, where the first excitonic peak has rather small shifts with changing environmental screening.

This linear scaling is significantly different from the $1/4$ scaling across different monolayer 2D materials~\cite{jiang2017scaling, choi2015linear}.
Physically, the scaling between $E_b$ and $E_g$ due to substrate screening has very different nature from the one of free-standing monolayer semiconductor. The environmental screening can be approximately described by classical electrostatic potential of dielectric interface~\cite{cho2018environmentally,waldecker2019rigid}, which gives a similar reduction on quasi-particle band gaps and exciton binding energies by
2D hydrogen model (linear scaling slope $\approx 1$).

On the other hand, the scaling between 2s binding energy and electronic gap is significantly smaller (0.65) than unity, which indicates the change of 2s exciton binding energy is a lot smaller than the electronic gap with increasing substrate screening. 
Therefore, the 2s exciton peak position is dominated by the change of electronic gap, which red shifts the spectra with increasing substrate screening (i.e. from vacuum to interfacing with SnSe$_2$ in Figure~\ref{figBSE1}). This stronger red shift of 2s exciton peak than 1s is also expected from the smaller reduction of 2s binding energy with increasing substrate screening in Figure~\ref{figNonrigid}. 

\subsection{Layer dependence of WS$_2$ optical spectra}

To further validate our method for substrate screenings on optical properties, we calculate the optical spectra from one to three-layer WS$_2$ with the GW/BSE method, then compare with recent experimental results~\cite{chernikov2014exciton}. The multi-layer calculations are performed with ``$\chi_{\text{eff}}$-sum" method introduced earlier~\cite{guo2020substrate}, which is computationally efficient and properly includes interlayer Coulomb interactions from first-principles.

\begin{figure}
    \includegraphics[width=0.45\textwidth]{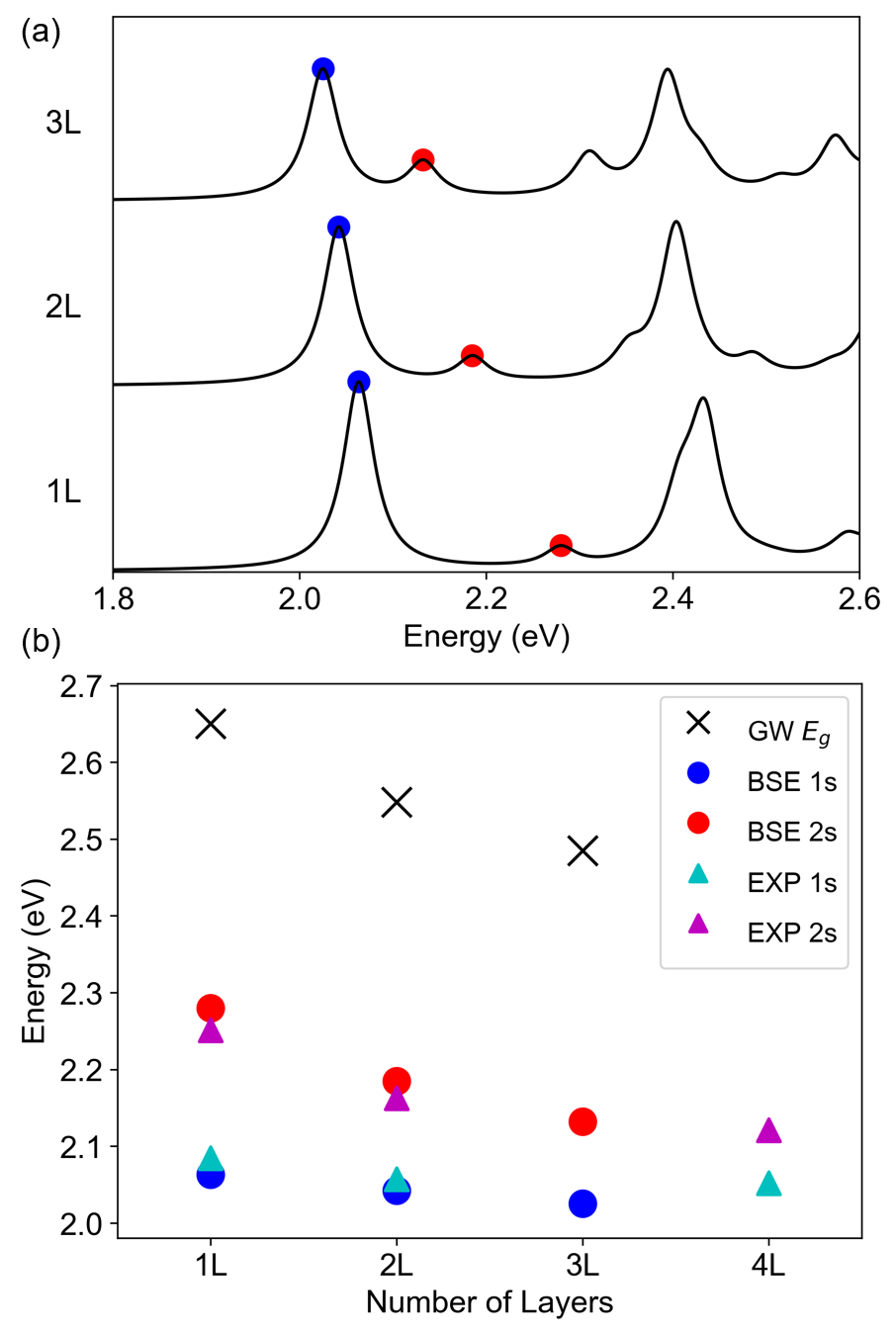}
    \centering

  \caption{\label{figBSE1} Layer dependence of optical properties of WS$_2$ a) calculated BSE absorption spectra for 1 layer(1L) $\to$ 3 layers (3L) WS$_2$ b) electronic gaps and optical excitation energies from GW and BSE, in comparison with experimental exciton energies~\cite{chernikov2014exciton}.}
\end{figure}

As shown in Figure~\ref{figBSE1}, we find the position of the first peak (1s, blue dot) is nearly unchanged (shifted within 20 meV) when increasing the number of layers, while the position of the second peak (2s, red dot) shifts over 100 meV. The calculated results with blue (1s) and red dots (2s) are compared with the experimental results (1s, blue triangle) and (2s, purple triangle). From 1L to 4L, the agreements between experiments and theory are nearly perfect, which validate the accuracy of our methods. 
Meanwhile, the calculated electronic gaps (black cross) are also shown in Figure~\ref{figBSE1}b, with a strong reduction as increasing the number of layers, in sharp contrast to the nearly unchanged optical gaps (1s exciton energies, red dots).

\subsection{Exciton lifetime in the presence of substrates}

\subsubsection{Zero temperature exciton lifetime}

Environmental screening due to substrates can also significantly modify the exciton lifetime $\tau$, which is a critical parameter that determines quantum efficiency.
The radiative rate $\gamma$ (inverse of lifetime $1/\tau$) based on the Fermi's Golden rule can be defined as follows~\cite{wu2019dimensionality, palummo2015exciton}:

\begin{align}
    \gamma(\vQ,\vq,\lambda) &= \frac{2\pi}{\hbar}\left|\mel{G,1_{\vq\lambda} }{\Hint}{S(\vQ),0}\right|^2\nonumber\\
    &\times \delta(E(\vQ)-\hbar cq_L), \label{eq:decay-general-Qqp}
\end{align}
where $G$ denotes the ground-state wavefunction,  $S(\vQ)$ the excited state,  $E(\vQ)$ the excitation energy,  $\vq$ photon wave-vector, $\lambda$ photon polarization direction and $\vQ$ exciton wave-vector. Then the radiative decay rate can be defined as the summation of each photon mode:  
\begin{align}
    \gamma(\vQ) &= \sum_{\vq\lambda=1,2}\gamma(\vQ,\vq,\lambda); \label{eq:decay-general-Q}
\end{align}
the corresponding radiative lifetime can be defined as the inverse of rate $\gamma(\vQ)$.
Furthermore, the radiative decay rate can be separated into two parts: 
\begin{align}
\gamma(\vQ) &= \gamma_0 Y(\vQ), \label{eq:rate-general-angle-separated}
\end{align}
where $\gamma_0$ is the exciton decay rate at $\vQ=0$ and $Y(\vQ)$ has exciton wave-vector dependence. 
Note that the $\vQ$ dependence will be only important to the exciton lifetime at finite temperature~\cite{wu2019dimensionality}.
Therefore the zero-temperature lifetime is simply $1/\gamma_0$.

Specifically for two-dimensional exciton lifetime at zero temperature, we have~\cite{wu2019dimensionality}
\begin{align}
  \gamma_0 = \frac{8\pi}{cA} \cdot \Omega \mu^2_S, 
\end{align}
where $\Omega$ is the exciton energy, $A$ is the unit cell area, and $\mu^2_S$ is the module square of dipole matrix elements~\cite{wu2019dimensionality}.

\begin{table}
\centering
\begin{tabular}{l l l l l}
\toprule
\text{Sub} & $\Omega/\text{eV}$ & $\mu^2_S \cdot \frac{8\pi}{V}/\text{eV}$ & $\tau_0/\text{fs}$\\
Vac & 5.34 & 0.788 & 30.9\\
hBN & 5.25 & 0.818 & 30.3\\
SnS$_2$ & 5.29 & 0.738 & 33.3\\
Gr & 5.26 & 0.706 & 35.0\\
SnSe$_2$ & 5.19 & 0.729 & 34.3\\
\end{tabular}
\caption{Monolayer hBN $1s$ exciton lifetime with different substrates at zero temperature, comparing with the free-standing one (Vac). $V$ is the volume of unit cell.}
\label{1slifetime}
\end{table}

\begin{table}
\centering
\begin{tabular}{l l l l l}
\toprule
\text{Sub} & $\Omega/\text{eV}$ & $\mu^2_S \cdot \frac{8\pi}{V}/\text{eV}$ & $\tau_0/\text{fs}$\\
Vac & 6.24 & 0.218 & 95.6\\
hBN & 6.04 & 0.201 & 107.1 \\
SnS$_2$ & 6.04 & 0.162 & 132.5\\
Gr & 5.94 & 0.120 & 182.8\\
SnSe$_2$ & 5.92 & 0.159 & 138.1\\
\end{tabular}
\caption{Monolayer hBN $2s$ exciton lifetime with different substrates at zero temperature, comparing with the free-standing one (Vac). $V$ is the volume of unit cell.}
\label{2slifetime}
\end{table}

The computed $1s$ and $2s$ exciton lifetimes of monolayer hBN on various substrates at zero temperature are shown in Table~\ref{1slifetime} and Table~\ref{2slifetime} respectively. 
The effect of substrate screening on $\tau_0$ comes from the quench of oscillator strength (or dipole moment $\mu^2_S$) and the red-shift of exciton energy, both of which increase the lifetime. In Table~\ref{1slifetime} and ~\ref{2slifetime}, $\mu^2_S$ are reduced by a similar amount between 1s and 2s excitons with increasing substrate screening; however, the relative proportion of reduction is much larger in 2s exciton due to its much weaker $\mu^2_S$ than 1s exciton. This results in 
stronger increase in 2s exciton lifetime  (i.e. increased by $30 \sim 80$ fs) in Table~\ref{2slifetime}. 
Instead, $1s$ exciton lifetime is rather insensitive to the substrate screening in Table~\ref{1slifetime}. 


\subsubsection{Finite temperature exciton lifetime}
The radiative exciton decay rate $\gamma(T)$ at finite temperature $T$ can be calculated by the thermal average of all accessable excitonic states as follows:
\begin{align}
    \gamma(T) &= \frac{\int \dd \vQ e^{-E(\vQ)/k_B T} \gamma(\vQ)}{Z}  \label{eq:rate-temperature-Z} \\
    Z &= \int \dd \vQ e^{-E(\vQ)/k_BT}, \label{partition}
\end{align}
where $E(\vQ)$ is the exciton energy dispersion as a function of exciton wave-vector $\vQ$.
Since a constant shift of $E(\vQ)$ does not change the expression of rate $\gamma$, we will use $E(\vQ) - E_0$ to replace $E(\vQ)$ in all later discussions, where $E_0=E(\vQ=0)$ is the lowest exciton  energy. As the integration of Eqs.~\ref{eq:rate-temperature-Z}-\ref{partition} requires the dispersion of exciton energy $E(\vQ)$, we use the  
effective mass approximation for exciton energy dispersion.

 



\begin{table}
\centering
\begin{tabular}{l l l l l}
\toprule
\text{Mat} & $\tau_0/\text{fs}$ & $\tau^{EM}_{RT}/\text{ps}$ & $\tau^{EXP}_{RT}/\text{ps}$\\
ML WS$_2$ & 334 & 923 & {806} ~\cite{yuan2015exciton}\\ 
\end{tabular}
\caption{\label{exp} Monolayer (ML) WS$_2$ 1s exciton radiative lifetime compares with experiment result. The zero temperature lifetime ($\tau_0$) is directly computed based on BSE exciton energy and dipole moments. The room temperature lifetime ($\tau^{EM}_{RT}$) is calculated with effective mass approximation with exciton effective mass $m_{exc} = 0.59$~\cite{shi2013quasiparticle}. The reference experiment result is obtained from {room temperature time-resolved photoluminescence (TRPL) spectroscopy}~\cite{yuan2015exciton} .}
\end{table}

\begin{table}
\centering
\begin{tabular}{l l l l}
\toprule
\text{Sub} & $\tau_0/\text{fs}$ & $\tau^{EM}_{RT}/\text{ps}$ \\
 Vac & 30.9 & 33.3 \\
hBN & 30.3 & 32.6 \\
SnS$_2$ & 33.3 & 35.8\\
Gr & 35.0 & 37.7\\
SnS$_2$ & 34.3 & 37.0\\
\end{tabular}
\caption{\label{ft} $\tau_0$ is the monolayer hBN ($1s$) exciton lifetime with diffient substrate at zero temperature (0K), while $\tau^{EM}_{RT}$ is room temperature lifetime with effective mass approximation at 300K compare with free-standing (Vac).}
\end{table}

First, we compute the room temperature (300K) exciton radiative lifetime of monolayer (ML) WS$_2$ to compare with experimental lifetime~\cite{yuan2015exciton}. The effective mass approximation for exciton dispersion is defined as $E(\vQ)=E(\vQ=0)+\hbar^2\QEx^2/2m_{exc}$, where $m_{exc}$ is the exciton effective mass.
$m_{exc}$ is chosen to be the summation of electron and hole effective mass, which was shown adequate for Wannier excitons ~\cite{mattis1984mass}. The electron ($m_e$) and hole ($m_h$) effective mass are from GW band structure results~\cite{shi2013quasiparticle} ($m_e = 0.27, m_h = 0.32$). Our calculated lifetime at room temperature is 923 ps for monolayer WS$_2$, in excellent agreement with experimental lifetime {806} ps~\cite{yuan2015exciton} as shown in Table~\ref{exp}.


We then apply the same methodology to compute the exciton radiative lifetime for
monolayer hBN with various substrates at finite temperature, as shown in Table~\ref{ft}. We find within the effective mass approximation, the room temperature exciton lifetime $\tau^{EM}_{RT}$ is much longer ($2 \sim 3$ orders) than the zero temperature lifetime $\tau_0$. To confirm the exciton lifetime of monolayer hBN with substrates in Table~\ref{ft} at finite temperature, future experimental work will be necessary. 

\section{Conclusion}
In this work we examined the substrate screening effects on excitonic excitation and recombination lifetime of 2D materials. We applied our previously developed effective polarizablility ($\chi_{\text{eff}}$-sum) method to efficiently calculate the electronic and optical spectra for arbitrary 2D interfaces with GW method and solving the BSE. We revealed the underlying mechanism of the non-rigid shifts of 1s and 2s peaks, i.e. why 2s red shifts much stronger than 1s in the presence of substrate screening. We explained this phenomenon through two steps: first, we showed a linear scaling (with a ratio of less than one) between 1s and 2s exciton binding energy both from our first-principle results and 2D Wannier exciton models; second, we presented the linear scaling between electronic gaps and exciton binding energies with a slope close to 1 for 1s and much smaller for 2s exciton while varying substrate screening. We further validated our method by reproducing the 1s and 2s exciton energy shift of WS$_2$ as a function of layer thickness observed experimentally. Finally we investigated the substrate effects on exciton lifetime and found the 2s exciton lifetime has a stronger dependence on substrates than 1s, due to the relative large change of exciton dipole moment.

\section*{Acknowledgements}

This work is supported by National Science Foundation under grant No. DMR-1760260 and DMR-1956015. This research used resources of the Center for Functional Nanomaterials, which is a US DOE Office of Science Facility, and the Scientific Data and Computing center, a component of the Computational Science Initiative, at Brookhaven National Laboratory under Contract No. DE-SC0012704, the lux supercomputer at UC Santa Cruz, funded by NSF MRI grant AST 1828315, the National Energy Research Scientific Computing Center (NERSC) a U.S. Department of Energy Office of Science User Facility operated under Contract No. DE-AC02-05CH11231, the Extreme Science and Engineering Discovery Environment (XSEDE) which is supported by National Science Foundation Grant No. ACI-1548562 \citep{xsede}.

\bibliographystyle{apsrev4-1}
\bibliography{ref}

\end{document}